# Experimental evidence of the ether-dragging hypothesis in GPS data

**Masanori Sato**

*Honda Electronics Co., Ltd.,*
*20 Oyamazuka, Oiwa-cho, Toyohashi, Aichi 441-3193, Japan*

**Abstract** In global positioning system (GPS) satellites, the earth-centered locally inertial (ECI) coordinate system is used for calculations. We cannot use other reference frames, for example one based on the solar system, in the calculation of GPS satellites, because if the relative velocities in the solar system are used, large periodic orbital deviations of reference time are calculated. Therefore, the ECI coordinate system is a stationary gravitational frame. This fact provides experimental evidence for the ether-dragging hypothesis in which the ether is assumed to be the permittivity of free space, $\varepsilon_0$, and the permeability of free space, $\mu_0$. This is interpreted using the analogy of an acoustic wave that is traveling in the atmosphere which is dragged by the gravity of the earth.

1. Introduction

   The global positioning system (GPS) is used in car navigation systems. The use of special relativity in GPS has been summarized by Ashby [1]. The GPS satellites orbit in a region of low gravity (20,000 km from ground level) at $v_G(t)$= 3.874 km/s. Therefore, the difference in gravitational potential between the ground and the location of the GPS satellites and the transverse Doppler shift effect of special relativity on the motion of the GPS satellites are considered. The transverse Doppler shift, or second-order Doppler shift, can be calculated by the Lorentz transformation of reference time. In GPS, the earth-centered locally inertial (ECI) coordinate system is used for the calculation of reference time. The ECI coordinate system is fixed to earth's center; thus, the earth is rotating as shown in **Fig. 1 (a)**. The GPS uses the relative velocities defined in the ECI coordinate system.

   Table 1 shows a summary of the GPS experiment from Ashby [1]. This discussion is carried out using the Lorentz transformation for the calculations.

Table 1 Summary of the GPS experiment from Ashby [1]

| Term | Time difference | Condition |
|---|---|---|
| Gravitational potential | 45.7 μs time gain every day | Height: 20,000 km from ground level |
| Velocity | 7.1 μs time delay every day | $v_G(t)$= 3.874 km/s |

Ether and explanations of gravitation were discussed from the 16$^{th}$ to 19$^{th}$ centuries by many of the greatest scientists, for example, Newton, Maxwell, and Lorentz.

From the equation of the phase velocity of electromagnetic wave, $c = \dfrac{1}{\sqrt{\varepsilon_0 \mu_0}}$, we assume that the ether is the permittivity of free space, $\varepsilon_0$, and the permeability of free space, $\mu_0$. ($\varepsilon_0$ is the electric permittivity of the ether in farads per meter; $\mu_0$ is the magnetic permeability of the ether in henrys per meter) Thereafter, the ether is dragged by the gravitational field of the earth. This classic idea was commonly disseminated and discussed in the days of the 19th century.

The aberration of light was observed by Bradley in 1725. He explained the aberration using Newton's particle property of photons, producing a simple illustration of a photon traveling in a straight line in the moving ether, without changing its direction. The aberration was considered to be one of the experimental results that show there is no ether-dragging around earth. This aberration is difficult to explain using the wave nature of the photon; however, it is easily explained using the particle nature of the photon. Therefore, the aberration does not rule out ether-dragging.

Michelson and Morley [2] denied the stationary ether. In the early 20$^{th}$ century, it was hypothesized that ether-dragging occurs on the ground level. To check this hypothesis experimentally, the Michelson-Morley experiment was carried out using massive lead blocks (one path of the interferometer was set between two lead blocks); there was no fringe shift [3]. Today, the GPS experiments show that there is no stationary ether at least up to 20,000 km, because the GPS works well at orbits of 20,000 km in height. If there is ether drift, it will be observed as an ether-wind more than 20,000 km from the ground level.

In 1933, Miller [4] reported experimental data that showed a slight seasonal and sidereal periodic fringe shift in the Michelson-Morley experiment. However, in 1955, his experimental results were re-evaluated and found to be thermal artifacts [5]. I believe that Miller's experimental results showed that the interferometer measurements are affected by the motion of the earth. I will discuss the ether-dragging hypothesis and the Michelson-Morley experiment in section 5.

In 1920, at the University of Leyden, Einstein described the ether in the theory of general relativity; he used "gravitational ether" in English translation [6]. It was a surprise for me to know Einstein emphasized the importance of the ether in the theory of general relativity. I consider that this was the first suggestion of applying the concept of ether to gravity with respect to the theory of general relativity.

In 1951, Dirac [7] referred to the ether in the context of his new electromagnetic theory. He suggested describing the ether from the viewpoint of quantum mechanics, that is, quantization of the ether. Dirac's suggestion is important to connect the theory of relativity and quantum mechanics.

Finally, using simple numerical calculations, we will be able to predict the periodic orbital reference time deviation in GPS satellites if the solar system is introduced [8]. However, in the GPS satellite experiments, such a large periodic orbital deviation that critically depends on the motion of the orbital plane of the GPS satellites in the solar system has not been detected [1]. Therefore, only the ECI coordinate system is correct. That is, the ECI coordinate system is a stationary reference frame. To explain this conclusion, one possible solution based on the ether-dragging hypothesis [9] is proposed where, the ether-dragging height is more than 20,000 km from ground level.

2. Consideration of the ECI coordinate system

The time of the GPS satellite is calculated based on the ECI coordinate system and works well. Why does the inertial system of the ECI coordinate system operate well? Twenty four satellites are launched on six orbits. It is supposed that the earth is moving at 30 km/s in the solar system, and the relative velocity of the earth in the solar system is set to $v_E$. The discussions are carried out in the solar system, thus, the velocities are relative to the solar system. There are two types of GPS orbits, one is the orbit P that is parallel to $v_E$, and the other is the orbit V that is perpendicular to $v_E$, as shown in **Fig. 1 (a)**.

Time dilation in the ECI coordinate system is calculated as follows:

$$t_G = \frac{t_0}{\sqrt{1-\left(\frac{\vec{v}_G}{c}\right)^2}} = \frac{t_0}{\sqrt{1-\left(\frac{3.874}{300,000}\right)^2}} \qquad (1)$$

$$\therefore \frac{t_G - t_0}{t_0} = \frac{1}{\sqrt{1-\left(\frac{v_G}{c}\right)^2}} - 1 = \frac{1}{\sqrt{1-\left(\frac{3.874}{300,000}\right)^2}} - 1 = 0.83377088 \times 10^{-10} \qquad (2)$$

where $t_0$ is the reference time, $t_G$ is that of the GPS satellite, $v_G(t) = 3.874$ km/s is the velocity of the GPS satellite defined in the ECI coordinate system, and $t_0$ is defined as the time on earth eliminating the gravitational effect and earth's rotational effect. A time delay is accumulated such that in 1 hour, the deviation of the car navigation system is roughly estimated to be $3,600 \times 0.833 \times 10^{-10} \times 20,000 \times 10^3 \, m \approx 6 \, m$. The accumulation of time dilation occurs because only the clock in the GPS satellite suffers the Lorentz transformation. This shows that the GPS is a very sensitive experimental setup for a time dilation measurement.

3. Calculation in the solar system

This section shows that the relative velocity defined in the solar system cannot be applied to the GPS calculations. A simple calculation shows that we can only use the relative velocity defined in the ECI coordinate system. The purpose of the numerical calculation is to show that there is a difference between the conditions in the gravitational potential of the earth and free space. Let us consider the GPS satellites moving in free space.

**Figure 1 (b)** shows the illustration of the calculation in free space. The GPS satellites orbit in the solar system, where there is no gravity and therefore no central force from the earth. The GPS satellites are connected by light rigid arms with the rotation center.

In the GPS experiment in free space, there is no gravity of the earth. Numerical calculations are carried out using the model in **Fig 1 (b)**, which shows periodic deviations. The difference between **Figs. 1 (a)** and **(b)** is the gravity. The model in **Fig. 1 (a)** is checked experimentally; however, that in **Fig. 1 (b)** is a thought experiment. Thus, the calculations are carried out in the solar system using the relative velocity of the earth $v_E$ and that of the GPS satellite $v_G(t)$.

The numerical calculations may defocus the discussion; therefore, I will summarize them in an appendix.

3.1 Orbits V and P

The orbit V is shown in **Fig. 2**, where the motion of the GPS satellite $v_G(t)$ is perpendicular to $v_E$, causing the orbit V to become a spiral trajectory in the solar system. The period of $v_G(t)$ is 12 hours. The time dilation is 7.2 μs per day. These calculations are shown in the appendix.

Next, the orbit P is considered. In **Fig. 3**, the summation of the velocity of the earth $v_E$ and the velocity of the GPS satellite $v_G(t)$ in the solar system is periodically changed every 12 hours. The reference time $t_G^P$ is calculated using equation (3) from the Lorentz transformation by setting $v_E$= 30 km/s and $v_G$= 3.874 km/s,

$$t_G^P = \frac{t_0}{\sqrt{1-\left(\frac{\vec{v}_E + \vec{v}_G(t)}{c}\right)^2}} \quad (3)$$

Equation (3) shows that there is a periodic deviation depending on the velocity $(\vec{v}_E + \vec{v}_G(t))^2$. The periodic derivation of the reference time $t_G^P$ is calculated as $\Delta t_G^P = \pm 1.3 \times 10^{-9}$; these calculations are shown in the appendix. The deviation $\Delta t_G^P$ is periodic, which causes a deviation in distance of around 0.28 km. However, the ECI coordinate system operates well by the GPS satellites, meaning no orbit-dependent periodic distance deviation is observed. The deviation of the reference time of the orbit P is similar to that of the orbit V. Thus, we cannot use the relative velocity defined in the solar system $(\vec{v}_E + \vec{v}_G(t))$.

3.2 Summary of the calculation

The calculations in the appendix are simple but numerous;; therefore, I will summarize them briefly here.

(1) The GPS is a very sensitive experimental setup to check the time dilation.

(2) GPS calculations cannot be carried out in the solar system using the relative velocity of the earth $v_E$ and that of the GPS satellite $v_G(t)$.

(3) A simple explanation for the ECI coordinate system is required. The reason for which the GPS works well in the ECI coordinate system but not in the solar system is discussed in Section 4 as a proposed solution.

(4) The difference between **Figs. 1 (a)** and **(b)** is gravity. The model in **Fig. 1 (a)** has been checked experimentally. Although the model shown in **Fig. 1 (b)** was derived from a thought experiment, the numerical calculation based on this model seems to be correct.

4. Proposed solution

Since the deviation of the reference time must be in agreement with the results of the ECI coordinate system, the calculation using the ECI coordinate system includes not only the effects of special relativity, but also of general relativity. However, a simple argument shows that it is almost impossible to calculate the GPS in the solar system.

One of the possible solutions is that the gravitational field of the earth is a stationary gravitational field. This is derived from the ether-dragging hypothesis in which the gravitational field of the earth drags the ether, where the ether is the permittivity of free space, $\varepsilon_0$, and the permeability of free space, $\mu_0$. The GPS works in the ether dragged by the gravitational field of the earth [9]. **Figure 4** shows the ether dragging model with the ECI coordinate system, the solar system, the galaxy, and the CMB are respectively in local stationary ethers. This is because each gravitational field drags the ethers around its gravitational field. The galaxy moves in the CMS at 700 km/s, the solar system moves in the galaxy at 230 km/s, and the ECI coordinate system moves in the solar system at 30 km/s, corresponding to points a to d in **Fig. 4** in local stationary states. The GPS satellite in the ECI coordinate system observes 4 km/s, but it does not detect the relative velocity in other coordinates. Thus, the GPS satellite is in the stationary state of the ECI coordinate system with relative velocity 4 km/s. If the GPS satellite leaves the ECI coordinate system from point a to point b, the local stationary state for the GPS satellite is changed to the solar system from the ECI coordinate system. The GPS satellite that moves parallel to the earth observes the velocity 30 km/s in the solar system. If we reach the gravitational field of Mars, we will be in another stationary gravitational state, namely the Mars-centered locally inertial coordinate system.

There may be another solution derived from the calculation using both special relativity and general relativity; however this calculation is very complicated. In this case, it becomes accelerated motion and the discussion using general relativity is required. The orbit P of the GPS satellite and the orbit of the earth seen from arbitrary inertial systems are shown in **Fig. 5**. The orbit P is a cycloid, so acceleration and deceleration are repeated. I have no idea how to carry out the calculation in an arbitrary reference frame. Therefore, I choose the ether-dragging hypothesis, which provides a simple calculation for the GPS experiments.

At the distance measurement of 20,000 km (this is a distance between the car and the GPS satellite), the time deviation $\Delta t_G^P = \pm 1.3 \times 10^{-9}$ causes a 26 mm deviation. ($20{,}000\,km \times 1.3 \times 10^{-9} = 26\,mm$) It is rather small; however, the deviation is accumulated for 6 hours. Let the value of the average deviation be half of the maximum deviation $\Delta t_G^P = \pm 1.3 \times 10^{-9}$. Thus, we obtain, multiplying a rough estimation of ½ the averaged deviation ½,

$$20{,}000\,km \times \frac{1}{2} \times 1.3 \times 10^{-9} \times 3{,}600 \times 6 = 0.28\,km. \tag{4}$$

This calculation shows that if the clock on the GPS satellite is left alone without an adjustment for 6 hours, the accumulated deviation of the distance is roughly 0.28 km. In the GPS, no such deviation is observed.

5. Discussion

The author thanks the reviewers for important comments on the earth's eccentricity as well as the suggestion that the discussion should not largely depend on the numerical calculations. From the viewpoint of the ether-dragging hypothesis, the Sagnac effects, the ECI coordinate system, the aberration, the Michelson-Morley experiments, and the analogy of an acoustic wave are discussed.

5.1 Earth's eccentricity

At perihelion the distance from the earth and the sun is 0.983 AU (astronomical unit: average distance between the earth and sun, $1.5 \times 10^{11}$ m), that of aphelion is 1.017 AU. Thus the velocity deviation is estimated to be 0.25 km/s, $(\vec{v}_E(t) = 30 \pm 0.25\,km/s)$, which causes about 1.6 mm of deviation at the measurement of 20,000 km,

$$20{,}000\,km \times \left(1 - \frac{\sqrt{1 - \left(\frac{29.75}{300{,}000}\right)^2}}{\sqrt{1 - \left(\frac{30}{300{,}000}\right)^2}}\right) = 1.6\,mm \quad . \tag{5}$$

The period of deviation is 12 months. The reference time deviation is not accumulated; this is because not only the clocks on the GPS satellite but also the clocks on earth simultaneously differ according to $v_E^2$. Therefore, there is no accumulation of the reference time deviations. Thus, the deviation relates to one time measurement of distance, for example, 20,000 km is a distance between the GPS and a car on earth; only 1.6 mm is calculated as the deviation. However, at the velocity of $v_G$= 3.874 km/s, only the clock on the GPS satellite obtains the deviation; thus, the deviation is accumulated for half the period of revolution as shown in equation (4). The accumulation causes a difference of between 0.28 km and 1.6 mm.

In the interferometer of the L1 band (1575.42 MHz, wavelength: 190 mm), relative accuracies of millimeters are reported [1]. This is the precise positioning analysis with carrier-phase measurements. However, at this stage, the resolution is 1% of the wavelength ($190\,mm \times 1/100 = 1.9\,mm$), thus the value of 1.6 mm is less than the technical limit of the measurement. However, the deviation of 1.6 mm is not observed for theoretical reasons. We cannot detect any annual deviations dependent on the velocity of the earth.

In this discussion, I assumed that the GPS is in the dragged ether, so we cannot detect any deviations. Even if advanced carrier-phase measurements with relative accuracies of sub millimeter are used, it is impossible to detect a 1.6 mm deviation that depends on the earth's eccentricity. The GPS experiments show that we cannot observe the reference time deviations represented by equations (3) and (5).

5.2 Sagnac effect

Sagnac effects reflect a changing distance between the light source and the observer caused by motion of the observer. If the observer moves during the flight time of the light, the distance between the light source and the observer is changed. For example, let the distance between the GPS satellite (signal source) and the observer on earth be 300,000 km (the distance that light travels at 1 second). On the equator, the speed of the ground is about 0.47 km/s. Thus the Sagnac effect is 0.47 km at the measurement of 300,000 km for the observer on the equator, meaning that the flight time of light of 1 second, the observer moves 0.47 km. According to Ashby [1], Sagnac effects are experimentally observed within a 2% deviation. Therefore, the deviation on the measurement of the Sagnac effect is calculated to be $0.47\,km \times 0.02 \div 300{,}000\,km = 3.13 \times 10^{-8}$. This value is equivalent to the deviation of the ether rotation in the ECI coordinate system. In other words, the ether is almost fixed to the ECI coordinate system.

Thus, the ether has two properties as shown in **Fig. 6**: 1) the ether is dragged with the earth, 2) the ether does not rotate with the earth; if the ether is fixed to the ECI coordinate system, the deviation of the rotation is roughly estimated to be less than $3.13 \times 10^{-8}$.

The angular frequency of the earth's rotation in the ECI coordinate system is $\omega_E = 7.2921151467 \times 10^{-5} \, rad \, s^{-1}$ [1], and the period is around 23 hours 56 minutes and 4.1 seconds. Thus, I assume that the earth rotates in the ether at the angular frequency $\omega_E$.

5.3 Proposed experiment of local positioning system

To explain the ether-dragging hypothesis, I describe other discussions of local positioning systems. The orbital center of the GPS satellites is the center of the earth, and the GPS satellites have symmetrical orbits and the velocity $v_G = 3.874$ km/s in the ECI coordinate system. Let us consider three geostationary satellites lying over the equator at 0 km, 6,400 km, and 12,800 km from the ground. The relative velocities in the ECI coordinate system are 0.47 km/s, 0.94 km/s, and 1.41 km/s, respectively. Therefore, three geostationary satellites observe time dilations that depend on their velocities.

The model in **Fig 1 (b)** can be checked around the earth in local positioning systems as shown in **Fig. 7**. If this local positioning system is set around the geostationary satellite, the relative velocities of the satellites defined in the ECI coordinate system have orbital deviations, and there are deviations of the reference time. Numerical calculations will show that the local positioning system needs rather complicated calculations to work well; in particular, we have to modify the calculation equations to use this system. If the proposed experiment is carried out, we may possibly obtain additional experimental evidence of the ether-dragging hypothesis.

**Figure 1 (b)** shows if there is no gravitational field of the earth, the ECI coordinate system is not defined. The origin of the ECI coordinate system is the earth; the ECI coordinate system is defined by the gravitational field of the earth. Therefore, the earth generates the ECI coordinate system.

5.4 Stationary state

What makes the ECI coordinate system and the stationary state? The answer is the gravitational field of the earth. The ECI coordinate system cannot be defined without the gravitational field of the earth.

**Figure 8** shows the earth and the local positioning system, the stationary state in the ECI coordinate system is also illustrated. The center satellite of the local positioning system is stationary in the ECI coordinate system, and we shall call it the ECI stationary satellite.

The origin that defines the ECI coordinate system is the earth; the origin of the stationary state in the ECI coordinate system is the earth. If this local positioning system is set away from the gravitational field of the earth, the ECI stationary satellite experiences the relative velocity of 30 km/s in the solar system.

The gravitational field of the earth generates the ECI coordinate system and the stationary state, in which time passes most rapidly. The time dilation occurs according to the relative velocities defined in the ECI coordinate system, which can be plausibly explained using the ether and ether-dragging hypotheses.

5.5 Aberration

The aberration is a counterargument against the ether-dragging hypothesis. If light is a wave in the ether, the light is dragged by the ether and the aberration cannot be observed on the earth. It is said that the aberration cannot be compatible with ether-dragging, but Bradley's explanation using Newton's particle model of light shows the compatibility of aberration and the ether-dragging hypothesis.

**Figure 9** shows aberration and ether-dragging. The wave-particle duality shows that a photon travels perpendicular to the wave front, which is also perpendicular in the dragged ether by the earth. The relative velocity between the stationary ether in the solar system and the dragged ether is 30 km/s. The boundary between the stationary ether and dragged ether is more than 20,000 km from ground level. Photons travel in straight lines in the solar system and dragged ether.

According to quantum mechanics, the phase velocity, c, of a photon is defined as $c=\omega/\kappa$ ($\omega$: frequency, $\kappa$: wave number); it is also the ratio of the energy $\varepsilon$ and momentum $\mu$ of the photon, $c=\varepsilon/\mu$. The energy and momentum are conserved to satisfy the constancy of the speed of light, c. Therefore a simple example of a photon traveling in a straight line in the moving ether without changing its direction is possible.

In the aberration, a photon is detected as a particle, and it is correct to use the particle properties of photons. Although it is impossible to explain the aberration by the wave property of photon, I do not consider that the aberration rules out ether-dragging.

5.6 Michelson-Morley experiment and GPS

The difference between the Michelson-Morley experiments and GPS experiments is the following: 1) the Michelson-Morley experiments are carried out in the rotating frame in the ECI coordinate system, 2) the GPS experiments are done in the ECI coordinate system. **Figure 6** shows the rotating frame (surface of the earth) and the ECI coordinate system. **Figure 9** shows that around the ground level on the equator, the relative velocity between the ground and dragged ether is 0.47 km/s. Thus, as described previously, the fringe shift observed in the Michelson-Morley experiments depends on the velocity of 0.47 km/s. It is rather difficult to observe the rotation of the earth with the Michelson interferometer, but the GPS can observe the rotation of the earth as a Sagnac effect at the sensitivity of 2 % [1], or a velocity of 9.4 m/s.

The GPS shows the isotropic constancy of the speed of light. The distance between the GPS satellite and the car is calculated using the time delay $t_D$ as $300,000\, km/s \times t_D(s)$, where the speed of light c=300,000 km/s is assumed to be an isotropic constant. Therefore, the fact that GPS works well is evidence for the isotropic constancy of the speed of light. The sensitivity of a direct one way measurement (from the GPS satellite to a car on earth) in GPS is $2 \times 10^4$ higher compared to the Michelson interferometer. Thus the null results are confirmed by the GPS.

Michelson and Morley [2] reported that the relative velocity of the earth and the ether is probably less than one sixth the earth orbital velocity (5 km/s) and certainly less than one fourth (7.5km/s). Not only Miller but also Michelson and others have reported the ether drifts below 10 km/s. Although, the deviations in the Michelson-Morley experimental results were considered as thermal artifacts [5], these deviations are critical evidence of the counterargument against ether-dragging. Thus, it is important to study the source of the deviations. In 1887, Michelson and Morley [2] assumed the earth's revolution in the solar system was 30 km/s. In 1933, Miller [4] assumed the solar system motion in the galaxy to be 200 km/s in addition to the earth's revolution in the solar system of 30 km/s. I consider that according to ether-dragging, null results are predicted; however, the experimental deviations observed by Miller and Michelson are large enough not to be negligible. I believe these experimental results of periodic deviations show that the interferometer measurements are affected by the rotation of the earth.

5.7 Analogy of acoustic wave

To make the discussion more clear, let us consider the analogy of an acoustic wave in the atmosphere, which is completely dragged by the gravity of the earth. Thus the motion of the earth in the solar system does not affect the speed of the acoustic wave. From this analogy, I derived the ether-dragging hypothesis.

The difference between the ether and the atmosphere is rotation. The atmosphere rotates synchronously with the earth with the angular frequency is $\omega_E = 7.2921151467 \times 10^{-5}\, rad\, s^{-1}$; on the other hand, the ether does not rotate (that is, $\omega_E = 0$).

At the beginning of the paper, from the analogy of an acoustic wave, I considered that the ether is fixed to the atmosphere and rotates synchronously with the earth. However, according to the GPS experiments, I think the ether is dragged but does not rotate; at this stage, I cannot explain the reason.

The GPS satellite is moving in the ECI coordinate system, in the solar system, in the galaxy, and in the CMB. However, the reference time of the GPS is only affected by the relative velocity defined in the ECI coordinate system. Numerical calculation shows that we cannot use the GPS in the solar system; this is one of the experimental pieces evidence for the ether-dragging hypothesis.

5.8 Summary of ether-dragging hypothesis

(1) The ether is the permittivity of free space, $\varepsilon_0$, and the permeability of free space, $\mu_0$.

(2) The ether is dragged by the gravitational field of the earth.

(3) The ether does not rotate with the earth; the ether is fixed to the ECI coordinate system.

(4) The ether-dragging hypothesis has been improving to be compatible with almost all the historical experiments as well as the GPS experiments.

(5) At this stage, we have to use the wave-particle duality in the explanation of the compatibility with the aberration.

6. Conclusion

In the GPS calculations, we can only use the ECI coordinate system; another reference frame, for example one based on the solar system, cannot be applied to GPS experiments. Therefore, the gravitational field of the earth is the stationary gravitational field, and the ECI coordinate system works well. This is the experimental evidence supporting the ether-dragging hypothesis. This hypothesis is interpreted using the analogy of an acoustic wave that is traveling in the atmosphere that is dragged by the gravity of the earth. I consider that we should reexamine the classical ether and ether-dragging hypothesis, which should have a compatibility with the theory of relativity, quantum mechanics, GPS experiments, and space physics.

References

1) N. Ashby, "Relativity in the Global Positioning System,"
   www.livingreviews.org/Articles/Volume6/2003-1ashby, (2003).

2) A. Michelson, and E. Morley, "On the Relative Motion of the Earth and the Luminiferous Ether," American Journal of Science, Third Series, **34**, 333, (1887),
   http://www.aip.org/history/gap/PDF/michelson.pdf.

3) G. Hammar, "The Velocity of Light within a Massive Enclosure," Physical Review **48**, 462, (1935).

4) D. Miller, "The Ether-Drift Experiment and the Determination of the Absolute Motion of the Earth", Reviews of Modern Physics, **5**, 203, (1933).

5) R. Shankland, S. McCuskey, F. Leone, and G. Kuerti, "New Analysis of the Interferometer Observations of Dayton C. Miller," Rev. Mod. Phys., **27**, 167, (1955).

6) A. Einstein, "Ether and the Theory of Relativity" (1920), republished in "Sidelights on Relativity," (Dover, NY, 1983).

7) P. Dirac "Is there an ether?" Nature, **168**, 906, (1951).

8) M. Sato, "Interpretation of special relativity as applied to earth-centered locally inertial

coordinate systems in Global Positioning System satellite experiments," arXiv:physics/0502007v4, (2006).

9) M. Sato, "A revisit of the papers on the theory of relativity: Reconsideration of the hypothesis of ether-dragging," arXiv:0704.1942v5, (2009).

**Appendix**: Numerical calculations in free space

(1) Orbit V

The orbit V is shown in **Fig. 2**, where a motion of the GPS satellite is perpendicular to $v_E$, thus causing the orbit V to become a spiral trajectory. The velocity of the GPS satellite in the ECI coordinate system is set to $v_G$. When the reference time of the reference frame at rest is set to $t_0$ and that of the rotation center is set to $t_E$ and is expressed with $t_0$, the Lorentz transformation is as follows:

$$t_E = \frac{t_0}{\sqrt{1-\left(\frac{\vec{v}_E}{c}\right)^2}} = \frac{t_0}{\sqrt{1-\left(\frac{30}{300,000}\right)^2}}$$

$$\therefore \frac{t_E - t_0}{t_0} = \frac{1}{\sqrt{1-\left(\frac{30}{300,000}\right)^2}} - 1 = 0.5 \times 10^{-8} \quad . \tag{A-1}$$

The motion of a GPS satellite is perpendicular to that of the rotation center such that $\vec{v}_E \perp \vec{v}_G$.

Because $\vec{v}_E \perp \vec{v}_G$, we can use the Pythagorean proposition to obtain the summation of $v_E$ and $v_G$ and the reference time $t_G^V$ of the GPS satellite by the following equation:

$$t_G^V = \frac{t_0}{\sqrt{1-\left(\frac{\vec{v}_E^2 + \vec{v}_G^2}{c^2}\right)}} = \frac{t_0}{\sqrt{1-\frac{30^2 + 3.874^2}{300,000^2}}}$$

$$\therefore \frac{t_G^V - t_0}{t_0} = \frac{1}{\sqrt{1-\frac{30^2 + 3.874^2}{300,000^2}}} - 1 = 5.08 \times 10^{-9} \quad . \tag{A-2}$$

Therefore, the proportion of the time delay over the rotation center of the GPS satellite is as follows:

$$\frac{t_G^V - t_E}{t_E} = \frac{t_G^V}{t_E} - 1 = \frac{\sqrt{1 - \frac{30^2}{300,000^2}}}{\sqrt{1 - \frac{30^2 + 3.874^2}{300,000^2}}} - 1 = 0.83377089 \times 10^{-10}. \quad \text{(A-3)}$$

The difference between equations (2) and (A-3) appears only after the 8th figure. Thus we obtain, $0.83377 \times 10^{-10} \times 60 \times 60 \times 24 = 7.2$ μs, and it becomes the delay of 7.2 μs per day (There is a difference with the experimental data of 7.1 μs, this difference critically depends on the value of the velocity $v_G$. In equation (A-2) and (A-3) we set $v_G$ =3.874 km/s).

(2) Orbit P

Next, the orbit P is considered. In **Fig. 3**, the summation of the velocity of the rotation center $v_E$ and the velocity of the GPS satellite in the solar system is periodically changed. A periodic derivation of the reference time $t_G^P$ is expected; for example, the reference time $t_G^P$ is calculated using equation (A-4) from the theory of special relativity by setting $v_E$= 30 km/s and $v_G$= 3.874 km/s.

$$t_G^P = \frac{t_0}{\sqrt{1 - \left(\frac{\vec{v}_E + \vec{v}_G(t)}{c}\right)^2}} \quad \text{(A-4)}$$

$$\therefore \frac{t_G^P - t_0}{t_0} = \frac{1}{\sqrt{1 - \left(\frac{\vec{v}_E + \vec{v}_G(t)}{c}\right)^2}} - 1. \quad \text{(A-5)}$$

1) At point A in **Fig. 3**, where $\vec{v}_E // \vec{v}_G$,

$$\frac{t_G^P - t_0}{t_0} = \frac{1}{\sqrt{1 - \left(\frac{30 + 3.874}{300,000}\right)^2}} - 1 = 6.42 \times 10^{-9}. \quad \text{(A-6)}$$

2) At point C in **Fig. 3**, where $\vec{v}_E // -\vec{v}_G$,

$$\therefore \frac{t_G^P - t_0}{t_0} = \frac{1}{\sqrt{1 - \left(\frac{30 - 3.874}{300,000}\right)^2}} - 1 = 3.75 \times 10^{-9} \quad . \tag{A-7}$$

3) At points B and D in **Fig. 3**, where $\vec{v}_E \perp \vec{v}_G$,

$$t_G^P = \frac{t_0}{\sqrt{1 - \frac{\vec{v}_E^2 + \vec{v}_G^2(t)}{c^2}}} \tag{A-8}$$

$$\therefore \frac{t_G^P - t_0}{t_0} = \frac{1}{\sqrt{1 - \frac{30^2 + 4^2}{300,000^2}}} - 1 = 5.08 \times 10^{-9} \quad . \tag{A-9}$$

The differences between equations (A-9) and (A-6), (A-7) is calculated as $\pm 1.3 \times 10^{-9}$, where $t_G^P$ is the reference time of the GPS satellite when $v_E$ and $v_G$ are parallel, and the deviation is calculated to be $\Delta t_G^P = \pm 1.3 \times 10^{-9}$. However, the deviation $\Delta t_G^P$ becomes one order larger compared to the value $0.8337 \times 10^{-10}$ obtained from equation (1). The deviation $\Delta t_G^P$ is sinusoidal; the periodic deviation is estimated to be more than 1 km. The ECI coordinate system operates well by the GPS satellites, and no periodic deviation is observed which depends on the orbits. The deviation of the reference time of the orbit P is similar to that of the orbit V.

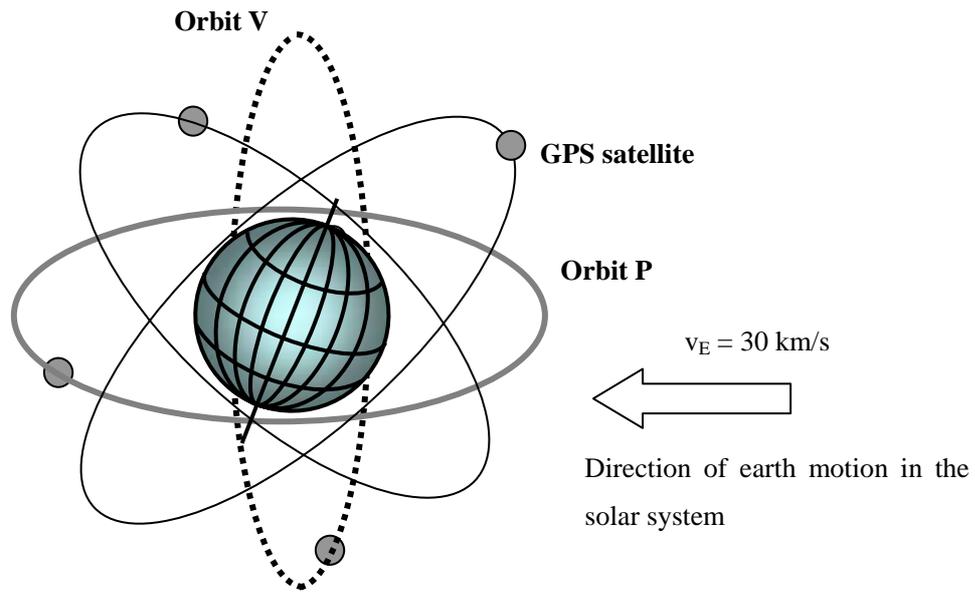

**Fig. 1 (a)** The earth-centered locally inertial (ECI) coordinate system. The GPS satellites and the earth motion in the solar system. Orbit P is parallel and orbit V is perpendicular to the direction of earth motion in the solar system, $v_E$= 30 km/s. The ECI coordinate system is fixed to earth's center; thus, the earth is rotating. The GPS satellites orbit at the relative velocity $v_G(t)$= 3.874 km/s defined in the ECI coordinate system. If the GPS satellites are seen from the solar system, their orbits are cycloid. However, the deviation of the GPS satellites is not observed. Time dilation only depend on the relative velocity $v_G(t)$= 3.874 km/s.

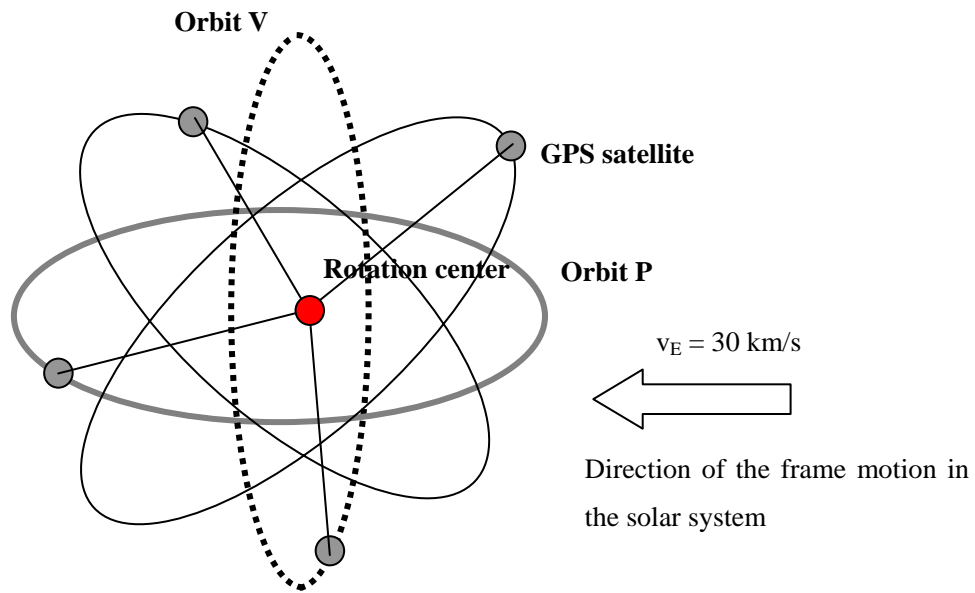

**Fig. 1 (b)** Thought experiment of the GPS in free space in the solar system: there is no gravity of the earth, GPS satellites and rotational center are combined by light rigid lodes. Numerical calculations are carried out using this model, which shows periodic deviations. The difference between **Figs. 1 (a)** and **(b)** is gravity. The model in **Fig. 1 (a)** is checked experimentally; however, that in **Fig. 1 (b)** is thought experiment, where the relative velocities are defined in the solar system.

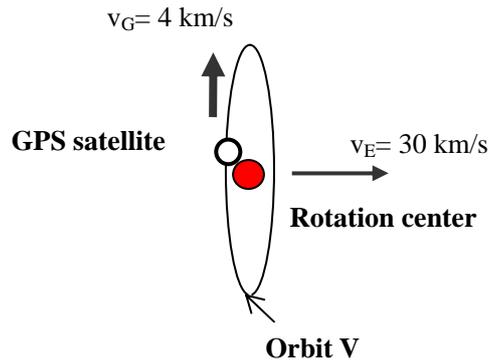

**Fig. 2** Orbit of satellite moving perpendicular to the direction of earth motion in the solar system. Orbit V: The orbital plane of the GPS satellite is perpendicular to the direction of the earth motion in the solar system.

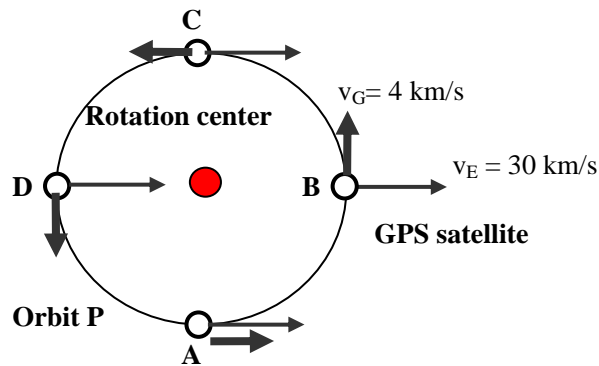

**Fig. 3** Orbit of satellite moving parallel to the direction of earth motion in the solar system (orbit P). The velocity of GPS satellite in orbit P has an orbital deviation. However, this illustration is not compatible with the GPS experimental data. There are no periodic orbital deviations.

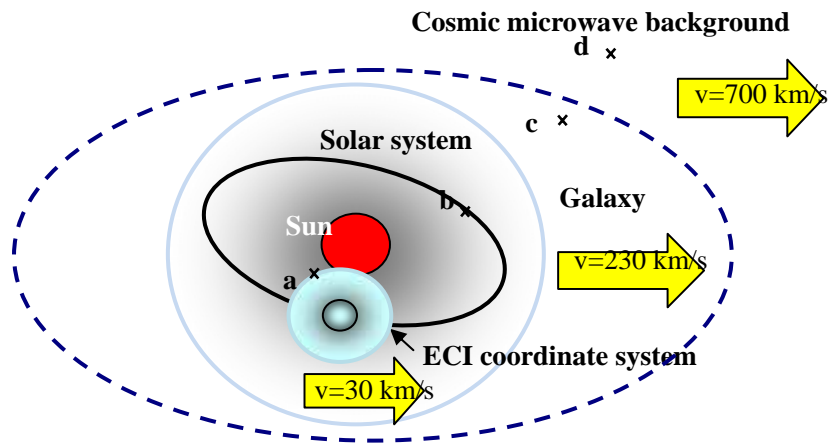

**Fig. 4** Ether dragging model: the ECI coordinate system, the solar system, and the CMB are respectively in the local stationary ethers. This is because each gravitational field drags the ethers around its gravitational field. There are many gravitational fields, thus, there are many local stationary states. Each points a to d are in local stationary states. If the GPS satellite leaves the ECI coordinate system, it will be into the local stationary state of the solar system.

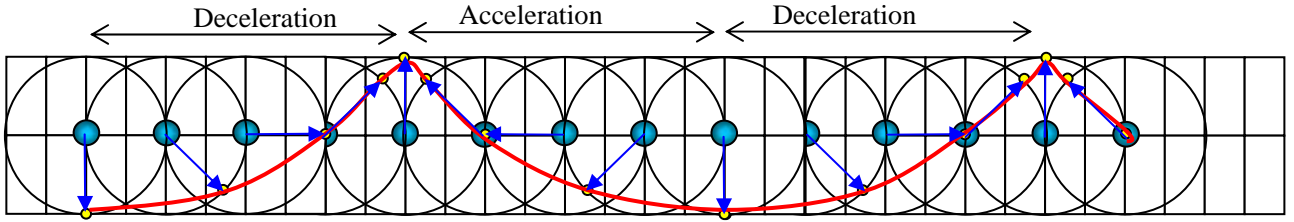

**Fig. 5** Traveling path of the GPS satellite of orbit P in an arbitrary reference frame. From an arbitrary reference frame, satellite motion is periodic, orbital accelerated motion of cycloid.

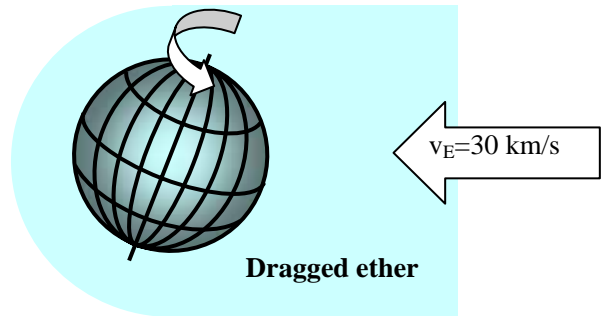

**Fig. 6** Ether-dragging model: 1) the ether is dragged with the earth, 2) the ether does not rotate with the earth; the ether is fixed to the ECI coordinate system.

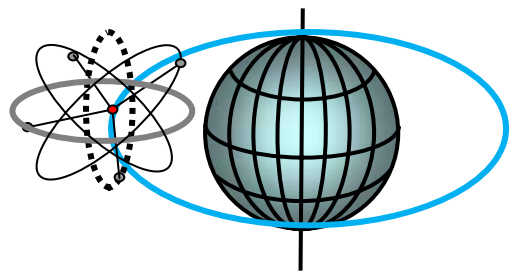

**Fig. 7** Local positioning system: this system orbits around the geostationary satellites lie over the equator. This proposed experiment shows that the gravitational field of the earth generates the ECI coordinate system.

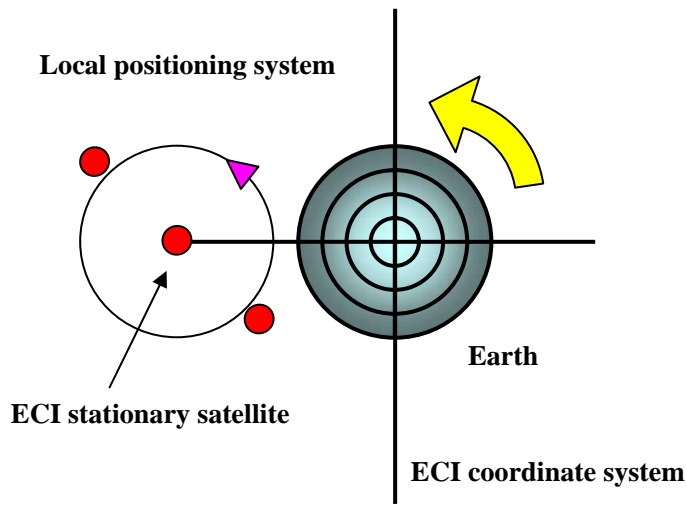

**Fig. 8** Stationary state in the ECI coordinate system: The center satellite of the local positioning system is stationary in the ECI coordinate system, let call it the ECI stationary satellite. The origin of the ECI stationary satellite is the earth. If this local positioning system is set out of the gravitational field of the earth, it suffers the relative velocity of 30 km/s in the solar system.

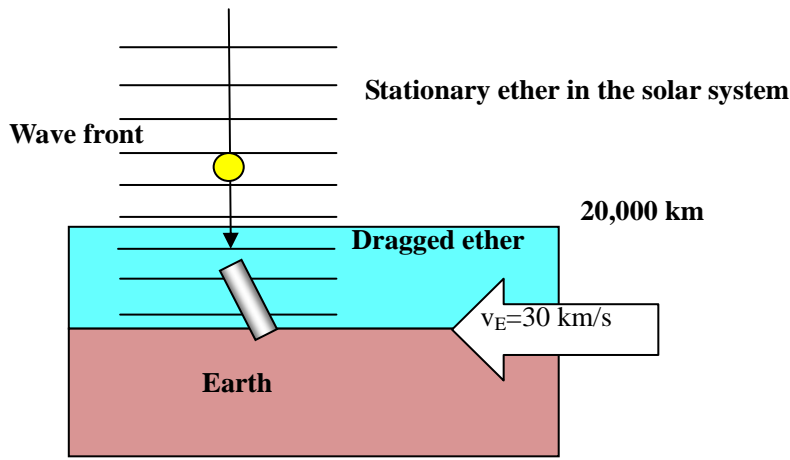

**Fig. 9** Aberration and ether-dragging: wave-particle duality shows that a photon travels perpendicular to the wave front which is also perpendicular in the dragged ether by the earth. Bradley's explanation using Newton's particle model of light shows the compatibility of aberration and ether-dragging hypothesis. The relative velocity between the dragged ether and the stationary ether in the solar system is 30 km/s. The boundary between the stationary ether and dragged ether is more than 20,000 km from ground level. Photons travel on the straight line in the solar system and dragged ether.